\documentclass[aps,prl,graphicx,9pt,twocolumn,superscriptaddress]{revtex4-2}
\usepackage[colorlinks,linkcolor=blue,urlcolor=blue,anchorcolor=blue,citecolor=blue]{hyperref}
\usepackage{amsmath}
\usepackage{graphicx}% Include figure files
\usepackage{dcolumn}% Align table columns on decimal point
\usepackage{mathrsfs}
\usepackage{amssymb}
\usepackage{amsfonts,multirow}
\usepackage{bm,changes}
\usepackage{color}
\usepackage{ulem}
\usepackage[thicklines]{cancel}

\begin{document}
% \title{Dynamical critical quantum sensing with a single parametrically-driven bosonic mode}
% \title{Dynamical critical quantum sensing with an interaction-free system}
% \title{Critical quantum sensing based on interaction-free Hamiltonian dynamics}
\title{Critical sensing with a single bosonic mode without boson-boson interactions}
\author{Ken Chen} 
\author{Jia-Hao L\"{u}}
\author{Xin Zhu}
\author{Hao-Long Zhang}
\author{Wen Ning}
\author{Zhen-Biao Yang}\email{zbyang@fzu.edu.cn}
\author{Shi-Biao Zheng}\email{t96034@fzu.edu.cn}
\address{Fujian Key Laboratory of Quantum Information and Quantum Optics, College of Physics and Information Engineering, Fuzhou University, Fuzhou, Fujian
 350108, China}

\begin{abstract}
    Critical phenomena of quantum systems are useful for enhancement of quantum sensing. However, experimental realizations of criticality enhancement have been confined to very few systems, 
    % owing to the stringent requirements, including adiabatic evolution, thermodynamical limit, and
    % %  precise control of interactions between quantum elements or particles. 
    % %  control of interaction Hamiltonian dynamics.
    % control of interactions between many particles or quasi-particles.
    owing to the stringent requirements, including the thermodynamical or scaling limit, and fine control of interacting quantum subsystems or particles.
     We here propose a simple critical quantum sensing scheme that requires
    %   none 
    neither
      of these conditions. The critical system is realized with a single parametrically-driven bosonic
    %   mode.
    mode involving many non-interacting bosons.
      We calculate the quantum Fisher information, and perform a simulation, which confirms the criticality-enabled enhancement. We further detail the response of one of the quadratures to the variation of the control parameter. The numerical results reveal that its inverted variance exhibits a diverging behavior at the critical point. Based on the presently available control techniques of parametric driving, we expect our scheme can be realized in different systems, e.g., ion traps and superconducting circuits. 
\end{abstract}
\maketitle

% \section{INTRODUCTION}
% \section{Introduction}
The quantum phase transition (QPT) \cite{sachdev_quantum_2011,10.1103/RevModPhys.69.315} is an interdisciplinary subject across a variety of physical branches, ranging from condensed physics, statistical mechanics to quantum optics. 
In addition to fundamental interest, critical phenomena near the QPT are valuable resources for quantum technological applications, among which quantum metrology is a paradigmatic example \cite{PhysRevA.93.022103,PhysRevA.96.013817,PRXQuantum.3.010354,10.1088/2058-9565/ac6ca5,PhysRevLett.120.150501,10.1364/OE.27.010482,PhysRevLett.128.173602,PhysRevLett.123.250401,10.1088/1402-4896/ab444c,PhysRevLett.126.200501,10.1038/s41467-021-26573-5}. 
At the critical point of a continuous QPT, the system's physical quantities, such as ground-state energy and excitation energy, exhibit divergent derivatives with respect to the control parameter. As a consequence, even a tiny variation of the control parameter can dramatically change the system's physical properties around the critical point. Such ultra-sensitive responses enable critical quantum systems to be used as quantum sensors with enhanced sensitivities. 
So far, critical quantum sensing has been demonstrated
in nuclear magnetic resonance 
% ensemble of two-spin systems 
two-spin ensembles
\cite{10.1038/s41534-021-00507-x} and non-equilibrium many-body systems of Rydberg atoms \cite{10.1038/s41567-022-01777-8}.
In recent years, many theoretical schemes have been proposed for criticality-enhanced quantum sensing in different systems, including the quantum Rabi model \cite{PhysRevLett.124.120504,PhysRevLett.126.010502,10.1007/s11433-022-2073-9}, the parametrically-driven Jaynes-Cummings model \cite{PhysRevA.106.062616}, and an Ising chain coupled to a cavity mode \cite{10.1088/1367-2630/13/5/053035}. 
All these schemes are based on the capability to precisely control freedom degrees of the corresponding composite light-matter system near the critical point, which is a challenging task in experiment.

Recently, a novel scheme was proposed to realize critical quantum sensing with a single bosonic mode \cite{10.1038/s41534-023-00690-z}. 
The QPT is driven by the combination of a 
% Kerr nonlinearity
Kerr-nonlinearity-induced photon-photon interaction
 and parametric driving \cite{grigoriou2023signatures}. 
This scenario is experimentally favorable as there is no need to control composite degrees of freedom. 
The parametric Kerr oscillator has been proposed to serve as a cat qubit for realizing fault-tolerant quantum computation \cite{10.1038/s41534-017-0019-1,doi:10.1126/sciadv.aay5901}, and was demonstrated in a recent superconducting circuit experiment \cite{10.1038/s41586-020-2587-z}. 
Despite these impressive advancements, it remains an outstanding experimental task to 
% realize such a critical system. 
well control this system near the critical point.

We here propose a simplified critical quantum sensing scheme. The critical enhancement is enabled with a single parametrically-driven bosonic mode, requiring neither the thermodynamical limit nor any nonlinear interaction. This simplification is important in view of experimental implementation of quantum metrological technologies. Considering the fact that parametric driving Hamiltonian can be well engineered in different systems, 
including optical systems \cite{PhysRevLett.57.2520}, ion traps \cite{PhysRevLett.76.1796,doi:10.1126/science.aaw2884,10.1038/s41567-021-01237-9} and superconducting circuits \cite{doi:10.1126/science.aaa2085,10.1038/s41586-020-2587-z}, 
our scheme would open promising prospects for critical quantum sensing with a single bosonic mode. We characterize the performance of the quantum sensing by quantum Fisher information (QFI) \cite{PhysRevLett.72.3439}. The numerical results show that the QFI tends to infinity at the critical point. The simulated inverted variance of 
% the field quadrature 
the squeezed quadrature 
exhibits a diverging behavior near the critical point, further evidencing the criticality-enhanced susceptibility.

% \section{The parametrically-driven bosonic mode}
We start by considering a parametrically-driven bosonic mode. In the interaction picture, the system Hamiltonian is given by
 ($\hbar=1$) \cite{scully_quantum_1997}
\begin{equation}
    \hat{H}=\omega \hat{a}^\dagger \hat{a}+\dfrac{\epsilon}{2}( \hat{a}^{\dagger 2}+ \hat{a}^2), \label{eq:H}
\end{equation}
where $ \hat{a}$ $(\hat{a}^\dagger)$ is the annihilation (creation)
operator for 
the bosonic mode,  
$\epsilon$ is the parametric driving strength, and $\omega$ denotes the detuning between the bosonic mode and the drive.
We note that this Hamiltonian corresponds to a parametric down conversion, which can be readily realized with optical systems \cite{PhysRevLett.57.2520}.

For simplicity, we define $g=\epsilon/\omega$ so that Eq. \hyperref[eq:H]{(1)} can be rewritten as $\hat{H}=\omega[ \hat{a}^\dagger \hat{a}+\dfrac{g}{2}( \hat{a}^{\dagger 2}+ \hat{a}^2)]$.
% \begin{equation}
%    H=\omega[ a^\dagger a+\dfrac{g}{2}( a^{\dagger 2}+ a^2)]. \label{eq:2}
% \end{equation}
To diagonalize the Hamiltonian, we perform the squeezing transformation
\begin{equation}
    \hat{H}_{\mathrm{np}}^\prime=\hat{S}^\dagger(r)\hat{H} \hat{S}(r), \label{eq:2}
\end{equation} 
where $\hat{S}(r)=e^{r( \hat{a}^{\dagger 2}- \hat{a}^2)/2}$ is the squeezing operator.
With the choice $r=\frac{1}{4}\mathrm{ln}\frac{1-g}{1+g}$,  the coefficient of the $(\hat{a}^{\dagger 2}+\hat{a}^2)$ term 
can be removed by the squeezing transformation, under which
 $\hat{H}$ is
diagonalized as $\hat{H}_{\mathrm{np}}^\prime=e_{\mathrm{np}} \hat{a}^\dagger \hat{a} +E_{\mathrm{np}}$
with the excitation energy $e_{\mathrm{np}}=\omega\sqrt{1-g^2}$
and the ground-state energy $E_{\mathrm{np}}=\frac{\omega}{2}\sqrt{1-g^2}-\frac{\omega}{2}$.
We note that the discrete eigenenergies exist only when $g \le 1$.
In this regime, the eigenstates and eigenenergies of $\hat{H}$ are
\begin{equation}
    \begin{aligned}
        |\phi_{\mathrm{np}}^n \rangle={}&\hat{S}(r)|n \rangle,  \label{eq:groundstates}\\
    \end{aligned}
\end{equation} 
and
\begin{equation}
    \begin{aligned}
        E_{\mathrm{np}}^n={}&ne_{\mathrm{np}}+E_{\mathrm{np}}, \label{eq:4}\\
    \end{aligned}
\end{equation}
respectively.

When $g<1$, the eigenspectrum is similar to that of a normal harmonic oscillator with an evenly energy spacing $e_{\mathrm{np}}$. The parametric driving transforms the Fock eigenstates  into squeezed Fock states. At the critical point $g=1$, the squeezing parameter becomes divergent, which serves as a signature of a QPT. 
For $g>1$, $e_{\mathrm{np}}$ becomes imaginary, which indicates that the energy gap vanishes and the spectrum becomes continuous. 
This phenomenon is something like the photon-blockage breaking phase transition, investigated in the driven Jaynes-Cummings model \cite{PhysRevX.5.031028,PhysRevX.7.011012}, where the detuning produced by the qubit-photon interaction cannot prevent continuous pumping of photons into the field when the driving strength is sufficiently large.
However, here the critical phenomenon is implemented with a single parametrically-driven bosonic mode. For the vibrational mode of a trapped ion, the parametric driving can be directly realized by applying an oscillating potential to the electrodes of the trap \cite{doi:10.1126/science.aaw2884}. For a microwave resonator, implementation of this driving would require additional systems \cite{doi:10.1126/science.aaa2085}, whose degrees of freedom, however, are adiabatically eliminated. Consequently, the control Hamiltonian involves neither a coupling between distinct freedom degrees, nor any interaction between bosons of the mode. This interaction-free characteristic of the control Hamiltonian distinguishes our criticality-enhanced quantum sensing protocol from previous ones \cite{PhysRevLett.124.120504,PhysRevLett.126.010502,10.1007/s11433-022-2073-9,PhysRevA.106.062616,10.1088/1367-2630/13/5/053035,10.1038/s41534-023-00690-z,10.1038/s41534-021-00507-x,10.1038/s41567-022-01777-8}, and represents a favorable virtue in light of experimental implementation.
We note the eigenenergies and eigenstates of the parametric down conversion Hamiltonian are similar to those of the Rabi Hamiltonian in the infinite qubit-photon frequency ratio \cite{PhysRevLett.115.180404}, although the present model does not involve the qubit, and hence is not subjected to the imperfection associated with the frequency ratio condition. This implies that the optimal measurement for realizing the critical quantum enhancement can be realized in a similar way as with the Rabi model \cite{PhysRevLett.124.120504,PhysRevLett.126.010502}.

\begin{figure}[htbp]
    \centering
    \includegraphics[width=\linewidth,page=1]{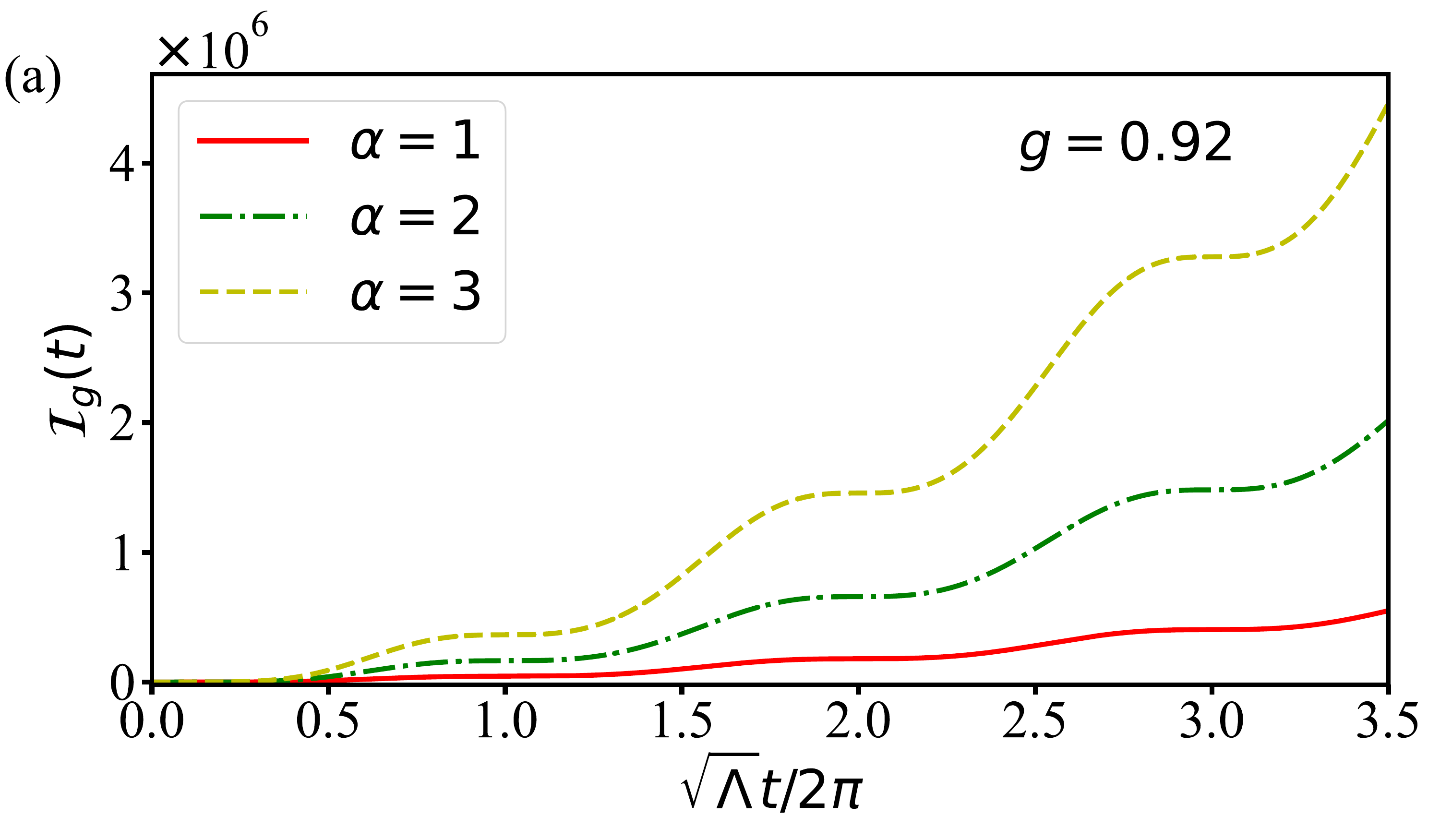}
    \includegraphics[width=\linewidth,page=1]{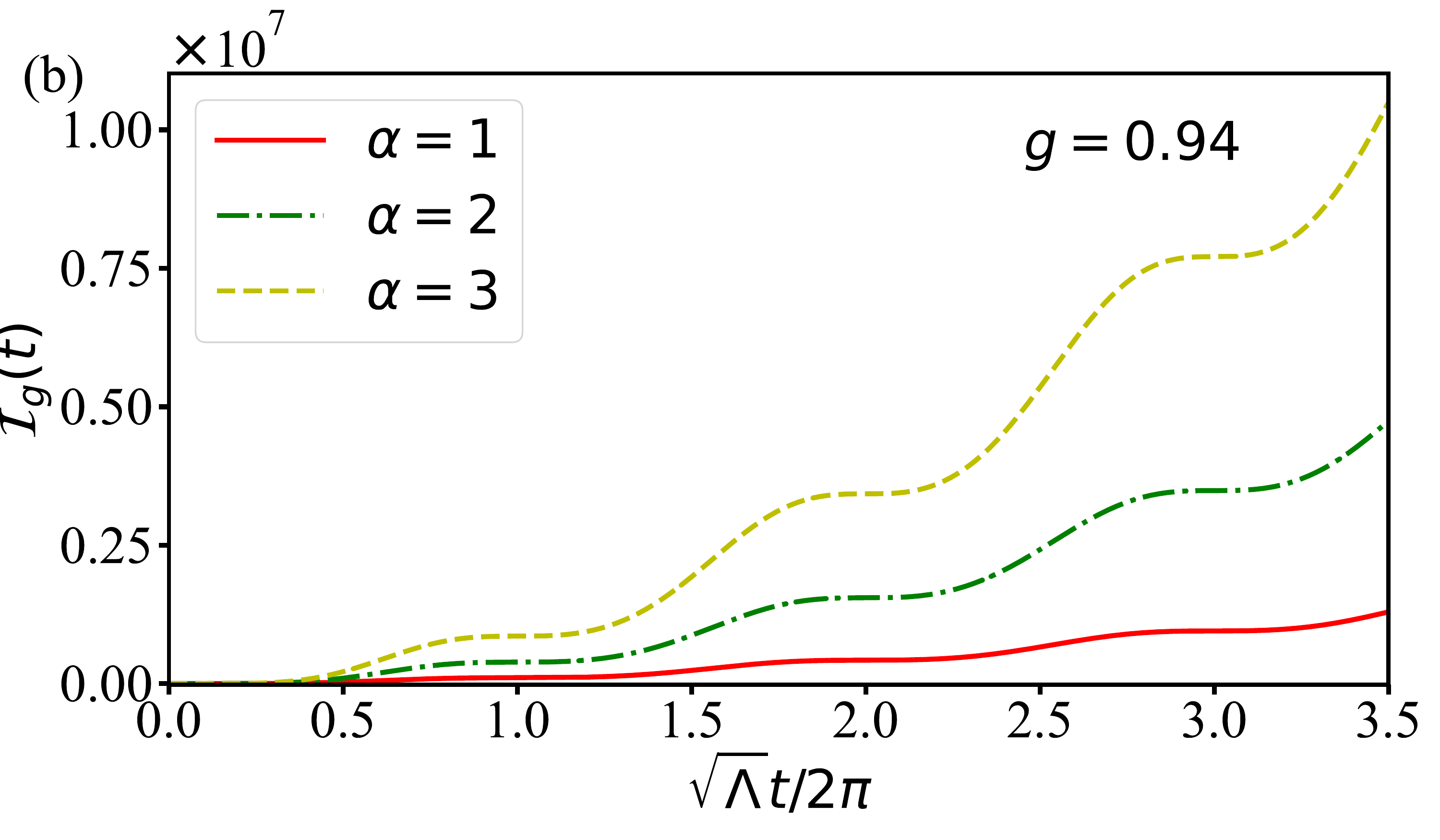}
    % {
    %     \includegraphics[width=0.4822\linewidth,page=1]{fig1a.pdf}
    % }
    % {
    %     \includegraphics[width=0.4822\linewidth,page=1]{fig1b.pdf}
    % }
    \includegraphics[width=\linewidth,page=1]{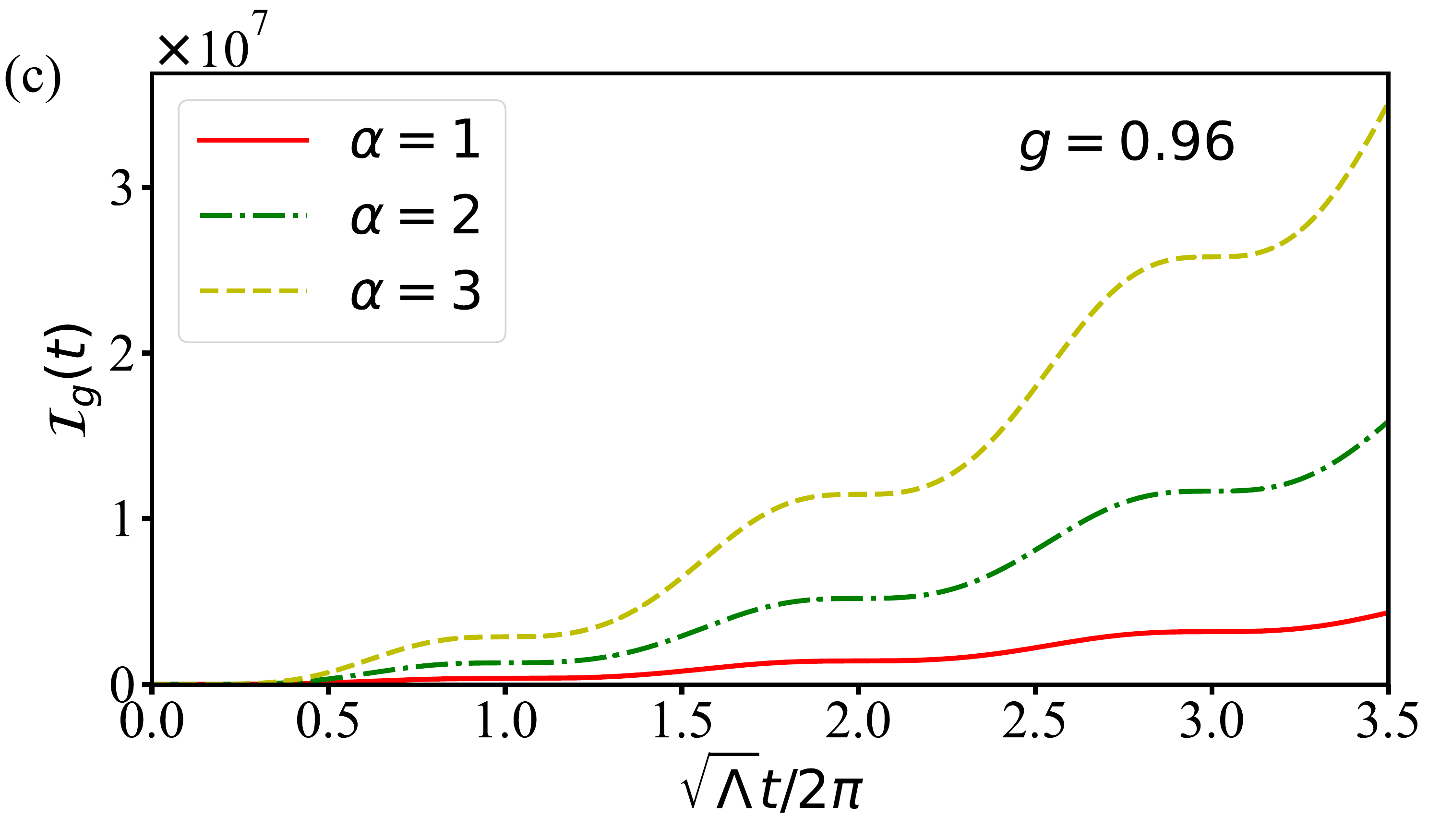}
    \caption{The QFI $\mathcal{I}_g(t)$ versus the evolution time $t$ for $g = 0.92$ (a), $0.94$ (b), and $0.96$ (c). Here $\alpha$ denotes the amplitude of the initial coherent state, and $g$ represents the dimensionless control parameter, which is defined as the ratio between the strength and detuning of the parametric driving.}    
    \label{Figure1}%文中引用该图片代号
\end{figure}

% \section{The QFI and a dynamic framework}
Note that the Hamiltonian of Eq. \hyperref[eq:H]{(1)}  satisfies a specific relation \cite{PhysRevLett.126.010502} so that the corresponding QFI with respect to the parameter $g$ is given by
\begin{equation}
    \begin{aligned}
        \mathcal{I}_{g}(t)\simeq 16\omega^6(1+g)^2 
        \frac{[{\rm sin}(\sqrt{\Lambda} t)-\sqrt{\Lambda} t]^2}{\Lambda^3}\delta[\hat{X}^2]_{{|\psi \rangle}},  \\
    \end{aligned}
\end{equation}
where 
\begin{equation}
    \Lambda=4\omega^2(1-g^2),
\end{equation}
and $\delta[\hat{X}^2]_{{|\psi \rangle}}$ represents the variance of $\hat{X}^2$ 
associated with the initial state $|\psi \rangle$, with $\hat{X}$ being the field quadrature $\hat{X}=(\hat{a}+\hat{a}^\dagger)/\sqrt{2}$. 
We set the bosonic mode to be initially in a coherent state $|\alpha\rangle$ with the amplitude $\alpha$ being a positive real number. Fig. \hyperref[Figure1]{1}(a), (b), and (c) show the QFI as a function of the evolution time $t$ for $g=0.92$, $0.94$, and $0.96$, respectively. In each subfigure, the solid, dot-dashed, and dashed lines denote the results for $\alpha=1$, 
$2$, and $3$, respectively. The results show the QFI increases with the increase of the amplitude of the initial coherent state.
It's not difficult to see that $\mathcal{I}_{g}$ tends to infinity as $g\to 1$ (i.e., $\Lambda \to 0$), 
which indicates the critical enhancement.

We first consider the problem of how to exploit the above-mentioned critical phenomenon to dynamically enhance susceptibility \cite{PhysRevLett.126.010502}. For the dynamical critical protocal, it is unnecessary to have the system adiabatically follow the ground state of the Hamiltonian, whose control parameter is slowly changed. The signal is encoded in the time-evolving state, governed by a time-independent Hamiltonian. For a specific initial state and with careful choice of the evolution time, the attainable precision can surpass the standard quantum limit \cite{RevModPhys.90.035006}, approaching the Heisenberg limit. 
% \textcolor{red}{We here assume that the system is initially in a coherent state $|\alpha\rangle$.}  
% \textcolor{red}{Preparation of such a superposition state of $|0\rangle$ and $|1\rangle$ needs the assistance of an ancilla qubit. In cavity QED experiments, the cavity mode can be prepared in the desired state by a swapping operation between the resonator and an effective two-level atom, which is initially in a superposition of its ground and excited states. After this operation, the atom-cavity interaction is switched off by tuning the atomic transition frequency far away from the cavity resonance, enabled by the Stark shift effect \cite{doi:10.1126/science.288.5473.2024}. }
After an evolving time $t$, 
the mean value and variance of the other quadrature $\hat{P}=i(\hat{a}^\dagger-\hat{a})/\sqrt{2}$ in the regime $g<1$ are
\begin{equation}
    \begin{aligned}
        % \langle\hat{P}\rangle_t={}&\sqrt{2}\alpha_i{\rm cos}(\omega t\sqrt{1-g^2})\\
        % {}&-\sqrt{2}\alpha_r\frac{(1+g)}{\sqrt{1-g^2}}{\rm sin}(\omega t\sqrt{1-g^2}), 
        \langle\hat{P}\rangle_t=-\sqrt{2}\alpha\frac{1+g}{\sqrt{1-g^2}}{\rm sin}(\omega t\sqrt{1-g^2}), 
    \end{aligned}
\end{equation} 
and
\begin{equation}
    (\Delta \hat{P})^2=\frac{1}{2} {\rm cos}^2(\omega t\sqrt{1-g^2})
    +\frac{1+g}{2(1-g)}{\rm sin}^2(\omega t\sqrt{1-g^2}), \label{eq:10}
\end{equation}
% \textcolor{red}{where $\alpha_r$ and $\alpha_i$ represent the real and imaginary component of $\alpha$, }
respectively.

With the quadrature $\hat{P}$ serving as the sensing indicator, the susceptibility, defined as $\chi_g(t)=\partial_g \langle \hat{P}\rangle_t$, becomes divergent in the vicinity of the critical point.
To clearly demonstrate the precision of the parameter estimation, we define the inverted variance $\mathcal{V}_g=\chi_g^2/(\Delta \hat{P})^2$, whose upper bound is imposed by the quantum Cram$\rm \acute{e}$r-Rao limit \cite{Harald0Mathematical, Rao1992}, i.e., 
$\mathcal{V}_g(t) \le \mathcal{I}_g(t)$.
The inverted variance $\mathcal{V}_g(t)$ is a periodic function of the evolution times $T_n=n\pi/(\sqrt{1-g^2}\omega)$ $(n \in \mathbb{Z}^\dagger)$ 
with the local maxima $\mathcal{V}_{g}(T_n)=4n^2 \pi^2 \alpha^2  \frac{g^2}{(1+g)(1-g)^3}$.

\begin{figure}[htbp]
    \centering
    \includegraphics[width=\linewidth]{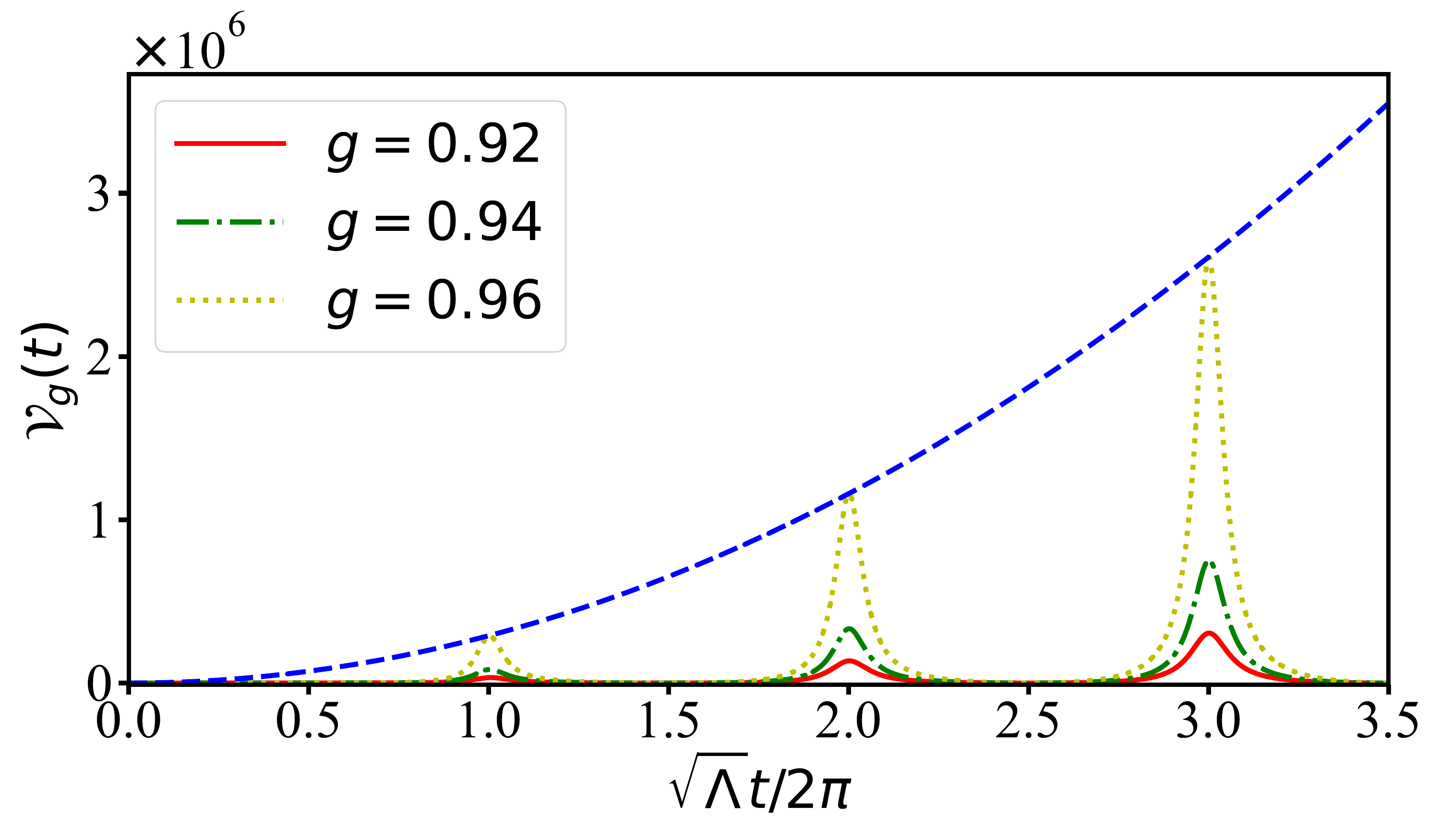}
    \caption{The inverted variance $\mathcal{V}_g(t)$ as a function of the evolution
    time $t$, based on the setting of three parameters $g = 0.92$, $0.94$, and $0.96$. The dashed line fits the local maximum of $\mathcal{V}_{g}(t)$ for $g=0.96$. 
    The result is calculated for the initial coherent state $|\alpha\rangle$  with $\alpha=1$.
    }
    \label{Figure2}%文中引用该图片代号
\end{figure}

Due to the periodic feature of inverted variance $\mathcal{V}_g(t)$, the performance of the dynamical protocol is sensitive to the choice of the evolution time. To
confirm this point, we plot
$\mathcal{V}_g(t)$ as functions of the time for different values of the control parameter $g$. The results are shown in Fig. \hyperref[Figure2]{2}. As expected, the inverted variance has a local maximum at 
% $t=2n\pi/\sqrt{\Lambda}$ $(n \in \mathbb{Z}^\dagger)$, 
$t=T_n$, 
% $\sqrt{\Lambda} t/(2\pi)=\sqrt{1-g^2}\omega t/\pi \in \mathbb{Z}^\dagger$, 
whose magnitude increases with $n$.
Compared with the dissipative-driving protocol proposed in Ref. \cite{10.1038/s41534-023-00690-z}, our scheme is based on the unitary dynamics and does not require the Kerr nonlinearity, which accounts for boson-boson interactions. This simplification is important for experimental implementation of the system, as well as for detection of the output signal as the Kerr nonlinearity would produce an unwanted phase evolution, distorting the produced quantum state \cite{PhysRevLett.115.137002}.

\begin{figure}[htbp]
    \centering
    \includegraphics[width=\linewidth]{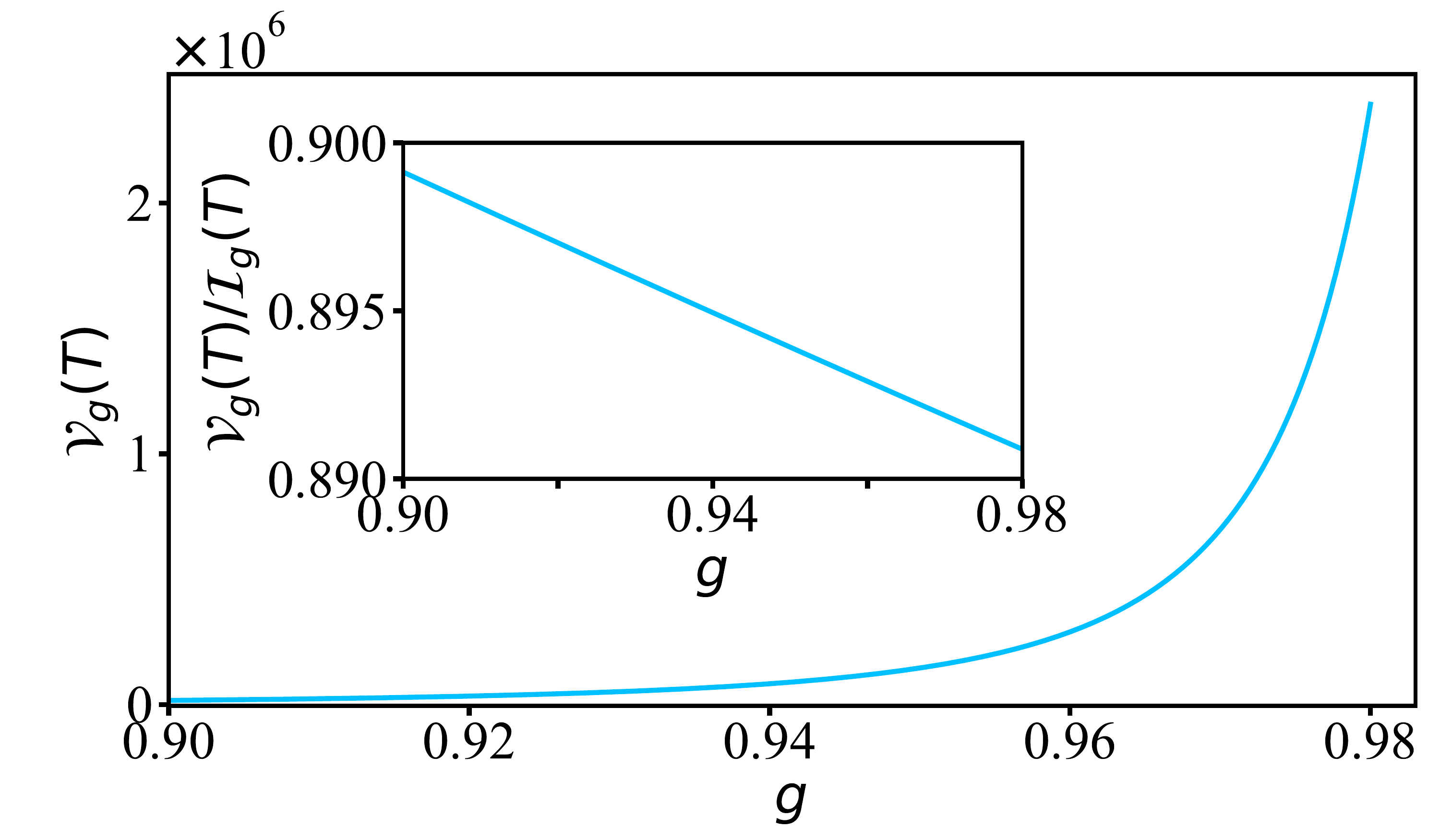}
    \caption{The inverted variance $\mathcal{V}_g(t)$ as a function of the
    parameter $g$ for an evolution time $T=2\pi/\sqrt{\Lambda}$. The inset shows the local maximum of the inverted
    variance $\mathcal{V}_g(T)$ arrives at
    the same order of $\mathcal{I}_g(T)$.
    The result is calculated for the initial coherent state $|\alpha\rangle$ with $\alpha=1$.
    }
    \label{Figure3}%文中引用该图片代号
\end{figure}

As shown in Fig. \hyperref[Figure2]{2}, the local maximum of the inverted variance nonlinearly increases with the corresponding measurement time. To quantitatively investigate the time scaling, we perform the power-law fitting $\mathcal{V}_{g}=Ct^{D}$ with the data of the local maxima. The dashed line in Fig. \hyperref[Figure2]{2} denotes the fitted result for $g=0.96$, with the parameter $D=2$. Fittings for other choices of $g$ near the critical point yield similar values of $D$. These results indicate that the critical protocol achieves the same precision scaling as conventional interferometric protocols with respect to the measurement time, imposed by the Heisenberg limit $\mathcal{V}_{g} \propto t^2$ \cite{PhysRevLett.124.120504}.

% \sout{In Fig. \hyperref[Figure2]{2}, the inverted variance $\mathcal{V}_g(t)$ is plotted versus the parameter $g$ for an evolution time $T=2\pi/\sqrt{\Lambda}$, exhibiting a divergent behavior at $g=1$.}
% \textcolor{red}{Fig. \hyperref[Figure3]{\textcolor{red}{3}} shows the inverted variance $\mathcal{V}_g(t)$ versus the parameter $g$ for an evolution time $T=2\pi/\sqrt{\Lambda}$. The amplitude of the initial coherent state $|\alpha\rangle$ is set to $\alpha=1$.}
% To a certain extent, we can see that the inverted variance $\mathcal{V}_g(T)$ and the QFI $\mathcal{I}_g(T)$ are on the same order of magnitude from the inset in \sout{Fig. \hyperref[Figure2]{2}} \textcolor{red}{Fig.} \hyperref[Figure3]{\textcolor{red}{3}}, which implies that the quadrature $\hat{P}$ serves as a good indicator for estimating the control parameter $g$.
% Besides, we find that as the evolution time $T_n$ increases, we can obtain larger $\mathcal{V}_g(T_n)$ value, 
% which corresponds to a higher precision. 
Fig. \hyperref[Figure3]{3} shows the inverted variance $\mathcal{V}_g(t)$ versus the parameter $g$ for an evolution time $T=2\pi/\sqrt{\Lambda}$. The amplitude of the initial coherent state $|\alpha\rangle$ is set to $\alpha=1$.
To a certain extent, we can see that the inverted variance $\mathcal{V}_g(T)$ and the QFI $\mathcal{I}_g(T)$ are on the same order of magnitude from the inset in Fig. \hyperref[Figure3]{3}, which implies that the quadrature $\hat{P}$ serves as a good indicator for estimating the control parameter $g$.
Besides, we find that as the evolution time $T_n$ increases, we can obtain larger $\mathcal{V}_g(T_n)$ value, 
which corresponds to a higher precision. 

We take our study one step further by investigating the performance of the adiabatic protocol, where the information about the physical quantity of interest is encoded in the adiabatically-changed ground state \cite{PhysRevLett.124.120504}. Below the critical point, the system has a unique ground state, which corresponds to a squeezed vacuum state $|\phi_{\mathrm{np}}^0\rangle$, defined in Eq. \hyperref[eq:groundstates]{(3)}, with the squeezing parameter depending on the control parameter $g$. The QFI with respect to the estimation of the physical quantity $A$ is $\mathcal{I}_A=4[\langle \partial_A \phi_{\mathrm{np}}^0 |\partial_A \phi_{\mathrm{np}}^0 \rangle+(\langle \partial_A \phi_{\mathrm{np}}^0 |\phi_{\mathrm{np}}^0\rangle)^2]$, where $\partial_A |\phi_{\mathrm{np}}^0 \rangle$ denotes the derivative of the ground state to $A$. Near the critical point, the QFI for estimating $\omega$ is dominantly contributed by the term
\begin{equation}
    \mathcal{I}_{\omega}=\frac{g^2}{2\omega^2(1+g)^2(1-g)^2}.  
\end{equation}
The attainable precision is bounded by the quantum Cram$\rm \acute{e}$r-Rao  limit: $\delta^2 \omega \ge \mathcal{I}_{\omega}^{-1}$, which sets the upper bound for the signal-to-noise ratio 
% $\mathcal{Q}_{\omega}\sim \frac{1}{2(1-g)^2}$. 
$\mathcal{Q}_{\omega}=\frac{g^2}{2(1+g)^2(1-g)^2}$. 

\begin{figure}[htbp]
    \centering
    \includegraphics[width=\linewidth,page=1]{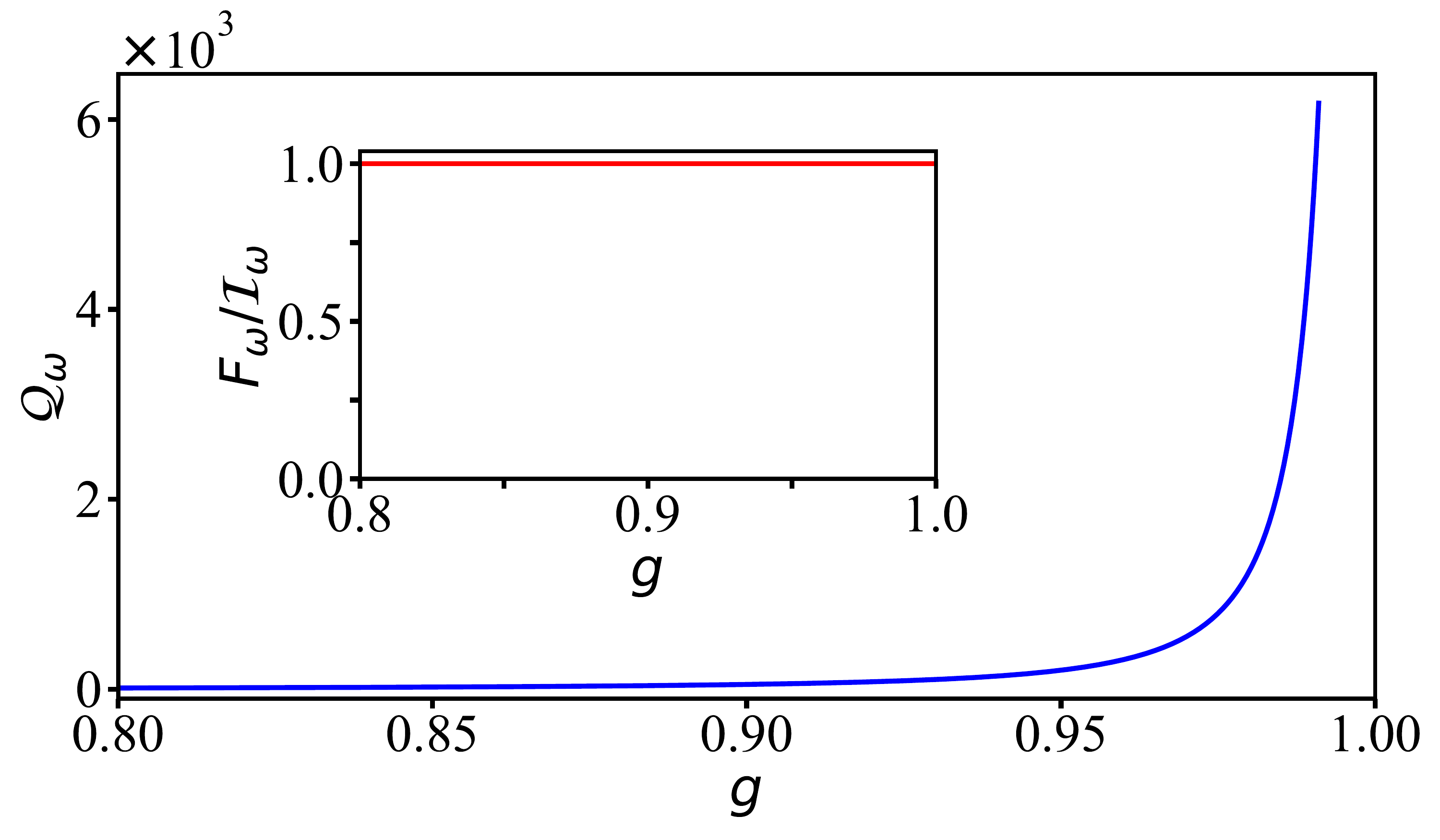}
    \caption{
    %     Signal-to-noise ratio $\mathcal{Q}_A$ versus $g$ for the estimation of $\epsilon$. Inset: the ratio FI/QFI versus $g$ for homodyne measurement of the $\hat{P}$ quadrature. In the normal phase $g < 1$, the optimal precision is reached for
    % all values of $g$.
    Signal-to-noise ratio $\mathcal{Q}_\omega$ versus $g$ for the estimation of $\omega$. Inset: the ratio $F_\omega/\mathcal{I}_\omega$ versus $g$. Here $\mathcal{I}_\omega$ denotes the QFI, while $F_\omega$ is the Fisher information associated with homodyne measurement of the $\hat{X}$ quadrature.
    }
    \label{figure4}%文中引用该图片代号
\end{figure}

To confirm the diverging behavior of $\mathcal{Q}_{\omega}$ near the critical point, we perform a numerical simulation and display the result in Fig. \hyperref[figure4]{4}. It is necessary to investigate whether or not the Cram$\rm \acute{e}$r-Rao  limit can be attained with respect to a practical measurement. We here verify this by calculating the Fisher information ($F_{\omega}$) of homodyne detection of the quadrature $\hat{X}=(\hat{a}^\dagger+\hat{a})/\sqrt{2}$, and comparing it to the QFI. 
This Fisher information is defined as $F_\omega=\mathop{\rm max}\limits_{\phi} \, \mathcal{P}_{\omega}[\hat{X}_\phi^2]$, where $\mathcal{P}_{\omega}[\hat{X}_\phi^2]=|\partial_\omega\langle \hat{X}_\phi^2 \rangle_\omega|^2/(\Delta \hat{X}_{\phi}^2)_\omega$, with $\hat{X}_\phi$ denoting the rotated quadrature operator $\hat{X}_\phi ={\rm cos}(\phi)\hat{X}+{\rm sin}(\phi)\hat{P}$. Substituting the squeezed vacuum $|\phi_{\mathrm{np}}^0 \rangle$ of Eq. \hyperref[eq:groundstates]{(3)} into the expressions of the expectation values, we obtain
\begin{equation}
    \mathcal{P}_{\omega}[\hat{X}_\phi^2]=f(\phi)\frac{g^2}{2\omega^2(1+g)^2(1-g)^2},
\end{equation}
where $f(\phi)=\frac{[{\rm cos}^2(\phi)e^{2r}-{\rm sin}^2(\phi)e^{-2r}]^2}{[{\rm cos}^2(\phi)e^{2r}+{\rm sin}^2(\phi)e^{-2r}]^2}$. As $\mathop{\rm max}\limits_{\phi} \, f(\phi)=1$, we have 
\begin{equation}
    F_\omega=\frac{g^2}{2\omega^2(1+g)^2(1-g)^2}.
\end{equation}
The ratio $F_\omega/\mathcal{I}_\omega$ versus $g$ is plotted in the inset of Fig. \hyperref[figure4]{4}, which demonstrates that this homodyne measurement saturates the Cram$\rm \acute{e}$r-Rao  bound for all values of $g$ in the regime $g<1$. We note that similar criticality-enhanced susceptibilities were previously revealed in protocols based on the superradiant phase transition of the quantum Rabi model \cite{PhysRevLett.124.120504,PhysRevLett.126.010502,10.1007/s11433-022-2073-9,PhysRevA.106.062616}, but where it is necessary to fine control the interaction between a photonic mode and a qubit. Furthermore, the superradiant phase transition occurs in the scaling limit, where the ratio between the frequencies of the qubit and light tends to infinity. In distinct contrast, our scheme is not subjected to these restrictions, thereby opening a promising perspective for criticality-enhanced quantum sensing.
In contrast with the dynamical implementation, the adiabatic protocol is robust against imperfect timing. To confirm this point, we compare the two cases: the control parameter $g$ is adiabatically increased to $0.98$ within times $T^\prime=100/\omega$ and $T^\prime=105/\omega$. In both cases, $g$ depends on $T^\prime$ as $g=kT^\prime/\sqrt{1+(kT^\prime)^2}$, where $k$ is the coefficient controlling the ramping velocity. For these two cases, the fidelities of the output states to the ideal squeezed vacuum state are $99.9936\%$ and $99.9934\%$, corresponding to signal-to-noise ratios $331$ and $373$, respectively.

% \section{Conclusion}
In conclusion, we have proposed a scheme for realizing quantum critical metrology with a single parametrically-driven bosonic mode without boson-boson interactions. Below the critical point, the system has a ladder-type eigenspectrum, underlain by squeezed Fock eigenstates. The diverging behavior of the squeezing parameter at the critical point enables realization of criticality-enhanced quantum sensing. We quantify the sensitivity with the QFI, which diverges at the critical point. Numerical simulations confirm the inverted variance of the squeezed quadrature exhibits a non-analytical behavior at the critical point. 
The Hamiltonian of Eq. \hyperref[eq:H]{(1)} can be synthesized in different systems, including ion traps \cite{PhysRevLett.76.1796,doi:10.1126/science.aaw2884,10.1038/s41567-021-01237-9} and superconducting circuits \cite{doi:10.1126/science.aaa2085,10.1038/s41586-020-2587-z}. In particular, both the degenerate parametric down conversion for engineering the Hamiltonian and homodyne detection for measuring the quadratures have been realized in optical cavity systems \cite{PhysRevLett.57.2520}. Therefore, our proposal is experimentally realizable with presently available techniques.

% \section*{Acknowledgments}
This work was supported by the National Natural
Science Foundation of China (Grants No. 12274080, and No. 11875108).

The authors declare no conflicts of interest.

Data underlying the results presented
in this paper are not publicly available at this time but may
be obtained from the authors upon reasonable request.

\bibliography{ref} 

\end{document}